\documentclass[aps,amsfonts,nofootinbib,superscriptaddress,twocolumn]{revtex4-1}

\usepackage{epsfig}
\usepackage{feynmp}
\usepackage{appendix}
\usepackage{graphicx}
\usepackage{dsfont}
\usepackage{amssymb}
\usepackage{tipa}
\usepackage{amsmath}
\usepackage{amsbsy}
\usepackage{color}
\usepackage{slashed}
\usepackage{bm}
\usepackage{bbm}
\usepackage{float}
\usepackage{ulem}
\usepackage{tikz}
\usepackage{pgfplots}
\usepackage{xparse}
\usepackage{upgreek}
\usepackage{stmaryrd }
\usepackage{caption}
\usepackage{subfigure}
\usepackage{soul}
\usepackage{dutchcal}
\usepackage{linguex}

\usetikzlibrary{positioning,arrows,patterns,automata}
\usetikzlibrary{decorations}
\usetikzlibrary{trees}
\usetikzlibrary{decorations.pathmorphing}
\usetikzlibrary{decorations.markings}
\usetikzlibrary{calc}

\newcommand{\ee}{\end{equation}}
\newcommand{\bb}{\begin{equation}}
\newcommand{\eqb}{\begin{eqnarray}}
\newcommand{\eqf}{\end{eqnarray}}

\tikzset{	aphoton/.style={decorate, decoration={snake}, draw=blue},
	photon/.style={decorate, decoration={snake}, draw=black},
	particle/.style={draw=black, postaction={decorate},
		decoration={markings,mark=at position .5 with {\arrow[draw=black]{>}}}},
	gluon/.style={decorate, draw=red,
		decoration={coil,amplitude=4pt, segment length=5pt}},
	vertex/.style={draw,shape=circle,fill=black,minimum size=3pt,inner sep=0pt},
}
\NewDocumentCommand\semiloop{O{black}mmmO{}O{above}}
{%
\draw[#1]     let    \p1     =    ($(#3)-(#2)$)     in    (#3)     arc
(#4:({#4+180}):({0.5*veclen(\x1,\y1)})node[midway, #6] {#5};)}

\begin{document}
\title{ Magnetic Seed and Cosmology as Quantum Hall Effect}
\author{H. Falomir}
\email{falomir@fisica.unlp.edu.ar}
\affiliation{Departamento de  F\'{\i}sica, Universidad Nacional  de La
  Plata, La Plata, Argentina}
  \author{J. Gamboa }
\email{jorge.gamboa@usach.cl}
\affiliation{Departamento de  F\'{\i}sica, Universidad de  Santiago de
  Chile, Casilla 307, Santiago, Chile}
  \author{P. Gondolo}
\email{paolo.gondolo@utah.edu.}
\affiliation{Departament  of Physics,  University of  Utah, Salt  Lake
  City, Utah, USA}
 \author{F. M\'endez  }
\email{fernando.mendez@usach.cl}
\affiliation{Departamento de  F\'{\i}sica, Universidad de  Santiago de
  Chile, Casilla 307, Santiago, Chile}

 \date{\today}

\begin{abstract}
  In the framework of a bimetric  model, we discuss a relation between
  the (modified) Friedmann equations and a mechanical system similar to the
  quantum Hall  effect problem.  Firstly, we  show how  these modified
  Friedmann equations are mapped  to an anisotropic two-dimensional charged
  harmonic oscillator  in the presence  of a constant  magnetic field,
  with  the frequencies  of the  oscillator  playing the  role of  the
  cosmological constants. This problem has two energy scales leading
  to  the  identification  of   two  different  regimes,  namely,  one
  dominated by the  cosmological constants, with exponential expansions
  for the  scale factors, and the other dominated by a  magnetic seed,
  which would be responsible for both a component of dark energy and a
  primordial magnetic  field.  The latter regime would be described by a (nonperturbative) mapping  between the
  cosmological evolution and the quantum Hall effect.
  \end{abstract}
\date{\today}
\maketitle

The   standard   description   of    the   Universe   rests   on   the
cosmological  principle, which  states  that,  on large  scales,
space-time  is   homogeneous  and  isotropic.  The   observations  are
consistent  with   this  hypothesis   for  distances  above   100  Mpc
\cite{trod,some}.  But   this  mathematical  idealization,   which  greatly
simplifies the  physical interpretation of the  model, has limitations
for lower scales. In particular, the formation of  structures can only
be understood  after the occurrence of  some gravitational instability
due      to      tiny      deviations     from      a      homogeneous
distribution~\cite{recent,varios}.

These departures from the cosmological  principle can be observed, for
example, in  the spectrum  of the  cosmic microwave  background (CMB),
which presents  temperature fluctuations of the  order of ${10}^{-5}$,
showing  that  corrections  to  classical cosmology  can  be
incorporated via perturbations~\cite{planck}.

However, one might wonder if there are other phenomena of cosmological
interest  that   might  require  a  non-perturbative   analysis.  This
possibility is particularly relevant since, in many fields of physics,
there are problems that are perturbative or non-perturbative depending
on the range of parameters one  is considering. As an example, one can
consider  a    gas  of charged  particles  subject  to  a
magnetic field  perpendicular to the  plane. If the magnetic  field is
strong enough,  the system  presents the quantum  Hall effect,  with a
Hamiltonian spectrum that can not be perturbatively obtained from that
of the free case.

This simple  example could  also be  translated into  the cosmological
regime by noting  that in the center of galaxies there are strong
magnetic fields which are observed through the Zeeman's splitting they
produce.  Although the  origin of these magnetic fields  is at present
unknown, the idea that a very small magnetic seed was formed in
an  early epoch  of the  universe evolution and
that, after a dynamo mechanism, the  field grew up to what is observed
today in galaxies   is widely  accepted \cite{Durrer:2013pga,Campanelli:2013mea, grasso,  holland,ratra}. Our present knowledge
does not  allow us to determine  when these magnetic seeds  were created,
but one  can speculate that  they might have  been originated in  the small
inhomogeneities existing before the recombination epoch.

{ Very probably the primordial magnetic fields did not produce any relevant effect after the recombination, but  these could be important in the first $100.000$-years and eventually to affect the big-bang nucleosynthesis, the dynamics of the phase transitions and even baryogenesis and leptogenesis \cite{electro1}.}

The magnetic seed must satisfy two consistency requirements. The first
one is that the coherence length is  not larger than about $10$ kpc, and the second
one is that  the field in the  magnetic seed must be  between $10^{-19}$
and $10^{-22}$ G.  In the analogous Hall system we discuss below,
the coherence  length corresponds to the magnetic depth $\ell_B$~\footnote{Here we use natural
units and $e=1$ \cite{hall}.}  \cite{cotas}, that is,
 \bb   \ell_B=    \frac{1}{\sqrt{B}}
<10~\mbox{kpc}. \label{bound}
\ee
The  second  condition  is
necessary  for the  stability  of the  dynamo mechanism  \cite{grasso, holland,ratra}.


A central issue  not solved so far is how to provide the cosmological standard model with a mechanism that incorporates a magnetic seed as a fundamental element \cite{Subramanian:2015lua}. Any possible answer to this question requires extra new ideas in a model that satisfies all constraints known so far and that incorporates the magnetic field as a central element. 

{{In this direction and using arguments coming from the formation of primordial magnetic fields \cite{grasso, holland,ratra} (we say for $t \sim 10^6$ years),  the mechanism  proposed here should work.}}

The  purpose  of  this  paper consists  in  investigating  the  possible
emergence  of magnetic  seeds  in a  model with  two  metrics with  an
effective interaction between them. This interaction can be considered as a relic
of  a causal  primordial connection  between sectors  in a  very early
epoch of the Universe. This problem  is considered in the context of a
simple   mechanical  system   that   nevertheless  reproduces   the
Friedmann's  equations of  the two  interacting sectors.  We emphasize
that  the important  issue  is not  only the  existence  of a  mapping
between  these apparently  unrelated systems  but also  that the  same
mechanism contributes to the production of dark energy.

{{ Interestingly, no matter how different the dark energy and magnetic seed scales might be
since in the present approach 
both are linked through a dynamical mechanism which (see Eq. ( \ref{i110})) allows to fix them in a rather independent 
way.}}

\medskip

In order to develop this idea let us consider the Lagrangian \footnote{The approach proposed here is valid for any number of patches, however  for simplicity in the presentation we will restrict ourselves to two of them.}
\bb
L= \frac{1}{2N} \left( {\dot x}_1^2 +{\dot x}_2^2\right) -\frac{N}{2}
\left( \omega_1^2 x_1^2+\omega_2^2 x_2^2 \right) - \frac{\theta}{2}
\left( x_1 {\dot x}_2 -{\dot x}_1 x_2 \right).
\label{lagran2}
\ee
Here  $x_1$ and $x_2$ are the dynamical variables, the  coefficients  $\omega_1,\omega_2$  and  $\theta$  are
constants, and  $N=N(t)$ is an  auxiliary variable that  transforms as
$N(t)  \rightarrow  t'(s) N(t(s))$  when  $t  \rightarrow t(s)$,  thus
ensuring the invariance of the action under time reparametrizations. This Lagrangian
yields the following Hamiltonian
\begin{eqnarray}
H    &=&    \frac{N}{2}    ~\bigg[p_1^2    +    p_2^2+\left(\omega_1^2
      +\frac{\theta^2}{4}\right)x_1^2+\left(\omega_2^2
         +\frac{\theta^2}{4}\right)x_2^2 +
\nonumber
\\
& &\theta \left(x_1 p_2 -x_2p_1\right) \bigg].
\label{i1}
\end{eqnarray}
This Hamiltonian  describes  an   anisotropic  two-dimensional  charged  harmonic
oscillator  with frequencies  $\omega_1$  and $\omega_2$,  interacting
with a constant magnetic field.

The Hamiltonian equations of motion  for  (\ref{i1}) are
\begin{equation}  {\dot x}_i  =  \left[ x_i,H\right],  ~~{\dot p}_i  =
  \left[ p_i, H\right],
\nonumber
\end{equation}
where $[~,~]$ is the Poisson  bracket, with the standard structure for
the canonical variables, that  is $\left[x_i,p_j\right] = \delta_{ij}$
and zero  for the remaining  brackets.  Alternatively, one  can define
the new  variables $\pi_i=  p_i -  \frac{\theta}{2} \epsilon_{ij}x_j$
and rewrite  the Hamiltonian $H=H(x_i,\pi_j)$  in order to  obtain the
equations of motion
\begin{equation}
{\dot  x}_i =  \left[  x_i,H\right], ~~{\dot  \pi}_i  = \left[  \pi_i,
  H\right],
\end{equation}
but with the following  Poisson brackets
\begin{equation}
\left[x_i,x_j\right] =0, \quad \left[x_i,\pi_j\right] = \delta_{ij},
\quad
\left[\pi_i,\pi_j\right] = \epsilon_{ij} \theta,
\label{j1}
\end{equation}
The equations of motion, once the momenta are eliminated, reduce to
\eqb
&&{\ddot x}_1 +\omega_1^2~ x_1 +\theta {\dot x}_2=0, \label{i01}
\\
&&{\ddot x}_2 +\omega_2^2 x_2 - \theta{\dot x}_1=0, \label{i02}
\\
&&{\dot x}_1^2+ {\dot x}_2^2+ \omega_1^2 x_1^2+\omega_2^2 x_2^2=0.
\label{i03}
\eqf
The  constraint   (\ref{i03})   is  a   consequence  of   time
reparametrization invariance  and, at the  end of the  derivation, the
gauge $N\equiv1$ has been chosen.

{Notice that this  constraint -- from the point of  view of the second
  order differential  equations (\ref{i01})  and (\ref{i02}) --  is in
  fact a relation between initial  conditions since the left hand side
  is  a  constant  of  the motion.   Indeed,  multiplying  (\ref{i01})  by
  ${\dot  x}_1$  and (\ref{i02})  by  ${\dot  x}_2$, and  adding  both
  equations, we immediately find that
\bb \frac{d}{dt}  \left[ {\dot
      x}_1^2 + {\dot x}_2^2 +\omega_1^2 x_1^2+ \omega_2^2 x_2^2
  \right] =0.
\ee
The  physical solutions  correspond to  those for  which the  constant
${\dot  x}_1^2 +  {\dot x}_2^2  +\omega_1^2 x_1^2+  \omega_2^2 x_2^2$
vanishes.  }

One  of  the  goals of  this  paper  is  to  point out  the  following
remarkable  mapping.   If we  redefine  the  variables $x_1,  x_2$  as
follows,
 \bb
x_1   =   \frac{2}{3}  a^{3/2}(t),   ~~~~~~~~~~~~~~~x_2=   \frac{2}{3}
b^{3/2}(t),
 \label{i4}
\ee
and replace  them in (\ref{i01})-(\ref{i02}), the  resulting equations
turn out to be
\eqb
 2\,  \frac{{\ddot   a}}{a}  +  \left(\frac{{\dot   a}}{a}\right)^2  +
 \frac{4}{3}\omega_1^2 &=& - { 2}\,\theta \sqrt{a\,b}~
 \frac{{\dot b}}{a^2},
 \label{i110}
\\
 2\,    \frac{{\ddot   b}}{b}+    \left(\frac{{\dot   b}}{b}\right)^2+
 \frac{4}{3}\omega_2^2 &=& { 2}\,\theta \sqrt{a\,b} \frac{ {\dot a}}{b^2},
  \label{i220}
\\
\label{i230}
a^3     \left[      \left(\frac{     {\dot      a}}{a}\right)^2     +
  \left(\frac{2}{3}\omega_1\right)^2\right] &=&
-       b^3\left[\left(\frac{\dot       b}{b}\right)^2+       {\left(
      \frac{2}{3}\omega_2\right)^2} \right].
\nonumber
\\
&&
\eqf
These equations are  identical to the Friedmann equations  for a cosmology  with two
metrics\footnote{The literature of cosmology  with two metrics is very
  extensive, see  for example \cite{bimetric} and   \cite{capo}.}  if  we identify their
respective cosmological constants $\Lambda_1$ and $\Lambda_2$ as
\begin{equation}\label{Lambdas}
    -\omega_1^2   \longleftrightarrow   \frac{3}{4}\Lambda_1,   \qquad
    -\omega_2^2 \longleftrightarrow \frac{3}{4}\Lambda_2.
\end{equation}

In  fact, Eqs.\  (\ref{i110})-(\ref{i230})  form a  coupled system  of
nonlinear second  order differential  equations for the  scale factors
$a(t)$    and    $b(t)$,   where    the    right    hand   sides    of
(\ref{i110})-(\ref{i220}) can  be considered as sources  of dark
  energy (See \cite{gondolo}  for a discussion on  a similar system and for string theory see \cite{string1}).
Moreover, from these equations one can read off the effective pressure
and  density  contributions induced  by  the  coupling between  scale
factors. Indeed,  expressing the Friedmann equations for the scale factor $a(t)$ in terms of the  pressure $p_b$ and energy density $\rho_b$ of an additional component of ``dark energy'', from
Eqs.\ (\ref{i110}) and (\ref{i230}) one obtains the equivalence
\eqb
8\pi G \,p_b&=& -{2}\,\theta \sqrt{a\,b}~
 \frac{{\dot b}}{a^2}, \nonumber
 \\
 \frac{8\pi G}{3} \rho_b &=& -\frac{1}{a^3}\left( { \frac{4}{9}\,\omega_2^2}\,
  b^3 +{\dot b}^2 b\right). \nonumber
 \eqf
This leads to the following equation of state for the effective component of dark energy,
\bb
\rho_b +\frac{{ 6} \pi G}{\theta^2} p_b^2 =
\frac{{ \Lambda_2 }\,}{{ 8} \pi G} \left(\frac{b}{a}\right)^3. \label{state}
\ee

For the case $\Lambda_2 =0$, the dark energy so
described  turns out to be a generalized Chaplygin gas \cite{Bento,Gorini,go1,go2,go3}. Notice that this is a non-perturbative result,  valid for any $\theta \neq 0$.

{Now we can use the mapping (\ref{i4}) to solve  the Friedmann equations.  Equations (\ref{i01})-(\ref{i02}) form a system of two coupled linear second order differential equations which consequently
have four linearly independent solutions. The latter have the form
\eqb
\label{frw2}
\left(
  \begin{array}{c}
    x_1(t) \\
    x_2(t) \\
  \end{array}
\right) = \left(
            \begin{array}{c}
             i \Omega \theta  \\
              \Omega^2 - {\omega_1}^2 \\
            \end{array}
          \right) e^{i \Omega t},
\eqf
where the frequency $\Omega$ takes one of the four values
\begin{eqnarray}
   \Omega_{\pm,\pm}  &=& \pm  \frac{1}{{ \sqrt{2}}}\bigg\{  \omega_1^2
                        +\omega_2^2 + \theta^2 \pm
   \nonumber
   \\
   &&\sqrt{\left[\omega_1^2+\omega_2^2   +   \theta^2   \right]^2   -4
      \omega_1^2 \omega_2^2}\bigg\}^{1/2}.
 \label{eigen2}
\end{eqnarray}
The general solution of  Eqs.\ (\ref{i01})-(\ref{i03}) is an arbitrary
linear  combination   of  these   four  functions   with  coefficients
${c_1,c_2,c_3,c_4}$.

To get  a solution of  our problem,  we must
also impose the constraint \eqref{i03}.}
Since the  left hand side of  \eqref{i03} is
  proportional to the Hamiltonian (written in terms of coordinates and
  velocities), it  is a constant  real symmetric
  quadratic form  in  $c_1,c_2,c_3,c_4$ (but
  not   positive   definite   for  the $\omega$'s    given   in   Eq.\
  \eqref{Lambdas}). The constrained solutions we are looking for correspond to the
  isotropic vectors of this  quadratic form.\footnote{The
    explicit   expression  of   this  quadratic   form  is   not  very
    enlightening,  so we  do  not  include it,  but  one can  convince
    oneself that it has a nontrivial isotropic subspace.}

The mapping  (\ref{i4})  allows us first,  to understand  the
evolution of the scale factors $  (a (t), b (t))$ under the previously
described  interaction through  the knowledge  of the  evolution of  a
mechanical system  of two  degrees of freedom $(x_1 (t),  x_2 (t))$,
and second,  to describe the  system by  means of the  Hamiltonian in
Eq.\ (\ref{i1}).

It is interesting  to note that the  limit $\omega_{1,2}^2 \rightarrow
0$ does not eliminate the  causal connection between metrics since, in
this case, the Hamiltonian (\ref{i1}) reduces to
\eqb
H&=&     \frac{N}{2}\left(     p_1^2    +p_2^2     +\frac{\theta^2}{4}
  \left(x_1^2+x_2^2\right)+\theta   \left(x_1   p_2  -x_2   p_1\right)
\right).
\nonumber
\end{eqnarray}
This can also be written as
\begin{equation}
H  =\frac{1}{2}\left[ {  \frac{\theta^2}{4}}  \left(  {\bar p}_1^2  +
    {\bar p}_2^2  \right) + \left({\bar x}_1^2  +{\bar x}_2^2\right) -
  \theta \left({\bar x}_1 {\bar p}_2 -{\bar x}_2 {\bar p}_1\right)
\right]. \label{nc1}
\end{equation}
Here   we   have  rescaled   variables   as   ${  \,x_i   =   2\,{\bar
    x_i}/\theta}$,  and  ${   p_i  =  \theta\,{\bar{p}}_i/2}$,  with
$i=1,2$,  and changed  $\theta\to-\theta$  to  obtain the  Hamiltonian
 considered  in \cite{nos} in  the context of  noncommutative
quantum mechanics.

Let us remark that the region where
 \bb
2|\omega_{1,2}| \ll |\theta| \label{reghall}
 \ee
is similar  to the strong  magnetic field  regime in the  quantum Hall
effect. In terms of cosmological constants this region corresponds to
 \bb
 \sqrt{3|\Lambda_{a,b}|}\ll |\theta|,
 \ee
 which can be called the {\it cosmological Hall regime}.

From  the   quantum  mechanical  point   of  view,  the   system described by the quantized Hamiltonian (\ref{i1}) is
 particularly interesting  because this implies replacing  the Poisson
 brackets (\ref{j1}) by the commutators (with $\hbar=1$)
   \eqb
 \left[{\hat x}_i,{\hat x}_j\right] &=&0, \quad \left[{\hat x}_i,{\hat
     \pi}_j\right]=i\,\delta_{ij}, \label{ent1}
 \\
 \left[{\hat     \pi}_i,{\hat    \pi}_j\right]     &=&i\,\epsilon_{ij}
 \theta,
 \label{ent}
 \eqf
where ${\hat \pi}_i = {\hat p}_i -\frac{\theta}{2} \epsilon_{ij} {\hat
  x}_j$,  with ${\hat  p}_i$  the canonical  momentum operator.  For
$\theta \neq 0$, the commutator  in Eq.\ (\ref{ent}) induces entangled
states for  $(x_1 (t),  x_2 (t))$ and, therefore,  for the
two metrics  of our  model, represented by  the scale  factors $(a(t),
b(t))$.

 The commutator (\ref{ent}) implements non-local communication between
 different spacetime regions, equivalently, entanglement states.

In addition, we note that in  the problem at hand  we have three
energy scales, $\sqrt {|\Lambda_a|}$, $\sqrt {|\Lambda_b|}$,  and $\sqrt{|\theta|}$. This allows us to
identify two regimes of cosmological interest, namely
\\
 (i) $\sqrt {|\Lambda_i|}\gg \sqrt  {|\theta|}$, which corresponds to a
 cosmological--constant dominated  era  in which  each metric  evolves
 independently with  no effective interaction, showing  an exponential
 behavior and  making the corresponding  side of Eq.\  \eqref{i230} to
 vanish;
  \\
 (ii) $\sqrt {|\Lambda_i|}\ll \sqrt  {|\theta|}$ which, by analogy, could
 be  interpreted  as the magnetic--seed  dominated era,  which
 eventually would  be responsible  for the  existence of  the magnetic
 fields in the universe.

The  quantum description  of these  two regimes,  which
could be relevant  in a very early epoch of  the Universe evolution,
is very different. In regime  (i) the system is formally described
(in  the  gauge  $N\equiv  1$)  by  the  Hamiltonian  operator  of  an
anisotropic harmonic oscillator.
On the other hand, in regime (ii) the equivalent mechanical system
is exactly a Landau problem, whose eigenstates are given by
 \begin{eqnarray}
 \psi_{n_+,n_-}   (x_1,x_2)   &=&  e^{\frac{\theta^2}{4}\left(   x_1^2
                              +x_2^2\right)}
                                  \left(\frac{\partial}{\partial  x_1}
                                  +  i  \frac{\partial}{\partial  x_2}
                                  \right)^{n_+} \times
 \nonumber
 \\
 & & \left(\frac{\partial}{\partial  x_1} - i \frac{\partial}{\partial
     x_2}     \right)^{n_-}     e^{-\frac{\theta^2}{4}\left(     x_1^2
     +x_2^2\right)}.
\nonumber
 \end{eqnarray}
The corresponding energy eigenvalues
 \bb
 \psi_{n_+,n_-}  = \theta \left(2 n_- +1\right)
\nonumber
 \ee
do  not  depend  of  $n_+$,   leading  to  an  infinitely  degenerate
Hamiltonian  spectrum.

This  regime  would  be  responsible  for
    inducing both a component  of dark energy  \cite{gondolo} and
    traces of  magnetic fields that  would subsequent grow.  In this
    sense, one might attribute both effects to a quantum origin of the
    Universe.

 We would  like to thank to  M. Henneaux, S. Mooij, M. Paranjape,  M. Plyushchay
 and  J.  C. Retamal  by  the  discussions.  This work  was  supported
 Dicyt and USA-1555 (J.G.),  Fondecyt-Chile 1140243 (F.M.). H.F.  thanks ANPCyT,
 CONICET  and  UNLP, Argentina,  for  partial  support through  grants
 PICT-2014-2304,  PIP  2015-2017 GI  -688CO  and  Proy. Nro.  11/X748,
 respectively and P. G was partially supported by NSF
Grant No. PHY-1720282 at the University of Utah.

\end{document}